\documentclass[a4paper,12pt]{article}
\usepackage{amsmath, epsfig,graphicx}

\usepackage{dpfloat}
\usepackage{wrapfig}
\usepackage{subfig}
\title{Statistical Microeconomics}
\author{Belal E. Baaquie
\\Department of Physics, National University of Singapore\\and\\Risk Management Institute, National University of Singapore\\ phybeb@nus.edu.sg}
\begin{document}
\maketitle
\begin{abstract}
A statistical generalization is made of microeconomics in the spirit of  going from classical to statistical mechanics. The price and quantity of every commodity\footnote{The term commodities is used for goods and services.} traded in the market, at each instant of time, is considered to be an \textit{independent random variable}: all prices and quantities are considered to be stochastic processes, with the observed market prices being a random sample of the stochastic prices. The dynamics of market prices is determined by an \textit{action functional} and, for concreteness, a specific model is proposed. The model can be calibrated from the unequal time correlation of the market commodity prices. A perturbation expansion for the correlation functions is defined in powers of the inverse of the total budget of the aggregate consumer and the propagator for the market prices is evaluated.
\end{abstract}
\section{Introduction}
The synthesis of economics and physics has given to rise of the new subject of Econophysics \cite{stanley}. Most of the studies in econophysics have been focused on the financial markets \cite{voit,bouchaud} and on financial instruments and their derivatives \cite{bebcup,bebcup2}. 

Microeconomics is one of the pillars of modern economic theory and studies the interaction of consumers and producers of commodities \cite{varian}, \cite{jehle}, \cite{green}. There is an increasing research in the application of statistical physics to economics \cite{chak2,gall,haven1,chak1} and this paper is a continuation of such studies.

Let quantity $\textbf{q}=(q_1,q_2,..,q_N)$, where $q_i>0$, be the quantity of a commodity labeled by $i$, with $i=1,2,..N$;  it can be kilograms of wheat or the number of automobiles. The commodity price vector is $\textbf{p}=(p_1,p_2,..,p_N)$, where $p_i>0$ is the price of a unit of the commodity; it can be dollars/kilograms or dollars/per automobile.  

One of the fundamental problems of microeconomics is to determine the dependence of quantities $\textbf{q}$ on the purchased at market prices $\textbf{p}$.

In most studies of microeconomics, at a given instant, the quantity and price of a commodity are taken to be a determinate quantity. Microeconomics  studies the (deterministic) equilibrium value of the quantities and prices of commodities as well their time evolution. A statistical  generalization is made of microeconomics by considering quantities ${q}_i(t)$ and price ${p}_i(t)$ to be independent random variables for each instant of time, namely \textit{stochastic variables}. 

A possible reason for prices to be random is that, similar to the price of equities, the prices of commodities incorporate all the market information and result in the traded prices. In absence of new information, any departures from the traded prices, hence, should be indeterminate, random and uncertain.  Furthermore, market prices are not in equilibrium, but rather have a (random) evolution in time $t$ that can have an overall drift reflecting market sentiment. Market prices may not contain all the market information and the source of randomness of market prices may have other explanations such as due to the existence of `sticky' prices \cite{sticky}.  

In statistical microeconomics, the supply $\mathcal{S}[\textbf{p}]$  and demand $\mathcal{D}[\textbf{p}]$  of commodities at market prices $\textbf{p}$ is the starting point for analyzing the behavior of the producers and consumers of commodities.  The competing tendency of demand and supply, namely demand increases when prices fall whereas supply increases when prices rise is reflected in the traded prices. In fact, in most microeconomics texts, the market commodity price is taken to be the value for which supply is equal to demand.

Supply and demand are inseparable, with one determining the other and vise versa. The view taken in this paper is that supply and demand are two facets of the same entity, namely a microeconomic \textit{potential function} $\mathcal{V}[\textbf{p}]$. Using the analogy from mechanics, a potential function $\mathcal{V}[\textbf{p}]$ is postulated that \textit{combines} supply and demand into a single entity and embodies the competing effects of both supply and demand. As will be discussed later, both the supply and demand functions are dimensionless and hence can be consistently added together. The potential is chosen to be the sum of supply and demand, namely 
\begin{eqnarray}
\label{defpot}
\mathcal{V}[\textbf{p}]=\mathcal{D}[\textbf{p}]+\mathcal{S}[\textbf{p}]
\end{eqnarray}
The potential function $\mathcal{V}[\textbf{p}]$, similar to mechanics,  drives the evolution of market prices. For the special case when the prices are constant (time independent) -- given by the constant prices $\textbf{p}_0=(p_{01},p_{02},..,p_{0N})$  -- the prices  \textit{minimize value} of the potential; namely that $\mathcal{V}[\textbf{p}_0]$ is a minimum of $\mathcal{V}[\textbf{p}]$.  In other words, in the framework of statistical microeconomics, stationary  prices are determined by the minimization of the microeconomic  potential, which replaces the standard microeconomic procedure of setting supply equal to demand \cite{varian}.

The full dynamics of market prices is determined by assigning a \textbf{joint probability distribution} for all possible evolutions of the stochastic market prices. In analogy with quantum mechanics and classical statistical mechanics, it is  \textit{postulated} that the probability of the stochastic evolution of market prices is proportional to the Boltzmann distribution, namely
\begin{eqnarray}
\label{boltzprop}
\text{Joint probability distribution}~~\propto~~\exp\{-\mathcal{A}[\textbf{p}]\}
\end{eqnarray}
where the action functional $\mathcal{A}[\textbf{p}]$ determines the likelihood of the evolution of all the different values taken by all the prices.

In analogy with mechanics, the action functional is taken to be the sum of the potential term $\mathcal{V}[p]$ with a \textit{kinetic term} $\mathcal{T}$, namely 
\begin{eqnarray}
\label{microaction}
\mathcal{A}[\textbf{p}]=\int_{-\infty}^{+\infty} dt\mathcal{L}(t)=\int_{-\infty}^{+\infty} dt \Big(\mathcal{T}[\textbf{p}(t)]+\mathcal{V}[\textbf{p}(t)]\Big)
\end{eqnarray}
with the Lagrangian given by
\begin{eqnarray}
\label{microlagrang}
&&\mathcal{L}(t) = \mathcal{T}[\textbf{p}(t)]+\mathcal{V}[\textbf{p}(t)]
\end{eqnarray}
The kinetic terms $\mathcal{T}[\textbf{p}(t)]$ contains the time derivatives of the prices and together with the potential function, determines the time dependence of the stochastic prices; in particular, $\exp\{-\mathcal{A}[\textbf{p}]\}$ determines the likelihood of the different random trajectories of the random prices.

Note that for all values of the prices $\mathcal{A}[\textbf{p}]>0$; the minimum value of $\mathcal{A}[\textbf{p}]$ has no significance, with the only requirement being that the minimum value is finite; by adding a constant, the minimum value of $\mathcal{A}[\textbf{p}]$ can always be taken to be zero.

To examine the specific characteristics of the statistical formulation of microeconomics, the total budget $m$ of a typical aggregate consumer is introduced as an expansion parameter. In particular, the correlation of the prices is studied as a perturbative expansion in a power series in $1/m$.  The perturbation expansion shows that the average prices of the model, to leading order in $1/m$, are equal to  the time independent stationary prices $\textbf{p}_0=(p_{01},p_{02},..,p_{0N})$ that minimize the potential. The series expansion of the unequal time price correlator -- in a power series in $1/m$ -- can be generated using the technique of Gaussian path integration

The model can be calibrated by comparing the model's unequal time correlation function with the empirical correlation of market commodity prices.

\section{The Utility Function}\label{sec:utility}
The utility function $\mathcal{U}$ is one of the fundamental concepts in microeconomics  and depends on the quantity of consumption vector $\textbf{q}=(q_1,q_2,..,q_N)$ of commodities, that is, $\mathcal{U}=\mathcal{U}[\textbf{q}]$. The utility function is a \textit{dimensionless real number} that quantifies  the utility of a commodity to the consumer, which is necessarily subjective. In all discussions in this paper,  the utility function $\mathcal{U}[\textbf{q}]$ refers to an `aggregate' consumer that reflects the norms of consumption of a given society -- and is not related to the subjective preferences of any specific individual.

A fundamental property of a utility function results from the intuitive expectation that a consumer gets more satisfaction by consuming greater quantities of a commodity; namely
\begin{eqnarray}
\label{fundproputility}
q'_i>q_i~~~ \text{if and only if}~~~ \mathcal{U}[q_1,q_2,..,q'_i,..q_N]>\mathcal{U}[q_1,q_2,..,q_i,..q_N]
\end{eqnarray}
Marginal utility is defined by the change in utility due to a change in the quantity consumed and is required to be positive, namely
\begin{eqnarray}
\label{marginalutility}
\text{Marginal utility}:~~\frac{\partial \mathcal{U}[\textbf{q}]}{\partial q_i} >0
\end{eqnarray}
The fact that marginal utility is a positive quantity  follows from Eq. \ref{fundproputility}.

The utility function is required to yield the so called \textit{diminishing marginal utility}, namely that consuming larger and larger quantities yields less and less marginal utility to the  consumer. Hence
\begin{eqnarray}
\label{dimmarginalutility}
\frac{\partial^2 \mathcal{U}[\textbf{q}]}{\partial q_i^2} <0~~:~~\text{Diminishing marginal utility}
\end{eqnarray}
There is a \textit{measurable consequence} of the  utility that a consumer derives from a commodity -- namely, the \textit{price} the consumer is willing to pay for the said commodity.  Let the total money available to the consumer be $m$; the consumer then has the following constraint on the quantities $q_i$ that are  consumed, namely that
\begin{eqnarray}
\label{pqconstraint}
\sum_{i=1}^Np_iq_i=m~~:~~\text{Budget constraint}
\end{eqnarray}

Given the finite budget of every consumer, the preferences of the consumer are reflected in the allocation of resources made by the consumer and results in different prices for different commodities. 

The utility function literally compares apples with oranges, since the consumer might prefer one commodity to another; to compare qualitatively different commodities, the utility function can only be a dimensionless function and of the dimensionless quantities $A_iq_i$, where $A_i$ has the inverse dimension of the quantity $q_i$; for example, if $q_i$ is the number of automobiles, the parameter $A_i$ has the dimension of per automobile; the numerical value of $A_i$ (per automobile) represents the importance of this commodity in the utility function. 

\section{The Demand Function} \label{sec:demand}
The demand function $\mathcal{D}[\textbf{p}]$ yields the prices $p_i$ for commodity $i$ that the consumer is willing to pay, given the budget constraint. 

Clearly, the demand function must be a decreasing function of prices, since, due to the consumers'  budget constraint, the higher the price of a commodity the smaller is the quantity that is bought by consumers. Hence
\begin{eqnarray}
\label{fundpropdemand}
p'_i>p_i~~~ \text{if and only if}~~~ \mathcal{D}[p_1,p_2,..,p'_i,..p_N]<\mathcal{D}[p_1,p_2,..,p_i,..p_N]
\end{eqnarray}

The demand function $\mathcal{D}[\textbf{p}]$ can be derived from the utility function and describes the empirical market demand of commodities. The aggregate consumer will consume quantities of commodities that maximize (optimize) the value of his (or her) utility function, subject to the budget constraint given in Eq. \ref{pqconstraint}. This yields
\begin{eqnarray}
\label{utilitydemandconst1}
\frac{\partial \mathcal{U}[\textbf{q}]}{\partial q_i}\Big{|}_{\textbf{q}=\bar{\textbf{q}}}=0\\
\label{utilitydemandconst2}
\text{Constraint}:~\sum_{i=1}^Np_iq_i=m
\end{eqnarray}
Simultaneously solving Equations \ref{utilitydemandconst1} and \ref{utilitydemandconst2} yields the value of $\bar{\textbf{q}}$ that maximizes the consumer's utility function for a given budget, namely
\begin{eqnarray}
\label{utidemandderiv}
\bar{\textbf{q}}=\bar{\textbf{q}}(\textbf{p},m)~~\Rightarrow ~~ \mathcal{D}[\textbf{p},m]=\mathcal{U}[\bar{\textbf{q}}(\textbf{p},m)]
\end{eqnarray}
Note that the demand function is dimensionless since the utility function is dimensionless. An example of a utility function and its corresponding demand function is analyzed in the Appendix; starting from a utility function, the demand is derived using the procedure of constrained optimization.

\subsection{Duality}
One can equivalently start from the demand function and  using the concept of \textbf{duality}. For a given demand function, together with the budget constraint, yields the following maximization problem
\begin{eqnarray}
\label{demandconst}
\frac{\partial \mathcal{D}[\textbf{p},m]}{\partial p_i}\Big{|}_{\textbf{p}=\bar{\textbf{p}}}=0\\
\label{demandconst2}
\text{Constraint}:~\sum_{i=1}^Np_iq_i=m
\end{eqnarray}
Simultaneously solving Equations \ref{demandconst} and \ref{demandconst2} yields the optimizing price $\bar{\textbf{p}}$
\begin{eqnarray}
\bar{\textbf{p}}=\bar{\textbf{p}}(\textbf{q},m)
\end{eqnarray}
and yields the utility function
\begin{eqnarray}
\label{demdndualuti}
\mathcal{U}[\textbf{q}]=\mathcal{D}[\bar{\textbf{p}}(\textbf{q},m),m]
\end{eqnarray}
One can view the demand function $\mathcal{D}[\textbf{p},m]$ as an \textit{indirect utility function}.

In deriving the utility function from the demand function, as given in Eq. \ref{demdndualuti}, the quantities of commodities $q_i$ was taken to be fixed and one maximized the demand function over all prices $p_i$. In contrast, in deriving the demand function from the utility function, as in Eq. \ref{utidemandderiv}, the commodity prices $p_i$ were taken to be given and quantities $q_i$ were varied to maximize utility.

\section{A Model Demand Function} \label{sec:demandutily}
Consider the following model for the demand function, namely
\begin{eqnarray}
\label{modeldemand}
\mathcal{D}[\textbf{p}]=\frac{m}{2}\sum_{i=1}^N\frac{d_i}{p_i^{a_i}} ~~;~~a_i,~d_i>0 
\end{eqnarray}
The coefficient $a_i$ is an index that characterizes the demand for a specific commodity; coefficients $d_i$ are determined by the relative importance of quantity $q_i$ in the demand for the total collection of $N$ commodities. All the coefficients $d_i>0$, that is, are positive since the demand function is positive, namely $\mathcal{D}[\textbf{p}]>0$. 

The form of demand function given in Eq. \ref{modeldemand} is quite realistic and, for example, has been used in an empirical study \cite{gas1} on the dependence of the demand of gasoline to its price; for the US market it was found that the index $a_\text{petrol}=0.075$ and the coefficient $md_\text{petrol}$ was taken to be a function of interest rates, inflation, per capita disposable income and so on.

The demand function $\mathcal{D}[\textbf{p}]$ is dimensionless and $m$ has the dimension of \$. 

The demand function clearly satisfies the condition stated in Eq. \ref{fundpropdemand}. The total demand is taken to be linearly proportional to the total budget $m$; this fulfills the requirement that if the consumer has no buying power there is no demand. The concept of `latent demand' that exists in the absence of buying power can be incorporated into the model by giving a time dependence to the budget constraint, namely $m=m(t)$, and is a feature that can be included in a more elaborate analysis of the model.

The utility function is obtained from the demand function using duality given in Eqs. \ref{demandconst} and \ref{demandconst2}. Using the method of the Lagrange multiplier, define an auxiliary function $B$ by
\begin{eqnarray*}
B= \sum_{i=1}^N\frac{d_i}{p_i^a} +\lambda(\sum_{i=1}^Np_iq_i-m)
\end{eqnarray*}
Minimizing $B$ with respect to both $p_i$ and $\lambda$, that is
\begin{eqnarray*}
\frac{\partial B}{\partial p_i}=0=\frac{\partial B}{\partial \lambda}
\end{eqnarray*}
yields
\begin{eqnarray}
\label{lambdacon}
\frac{a_id_i}{p_i^{a_i}}=\lambda p_iq_i~~;~~\sum_{i=1}^Np_iq_i=m
\end{eqnarray}
From above equations
\begin{eqnarray*}
\lambda=\frac{1}{m} \sum_{i=1}^N\frac{a_id_i}{p_i^{a_i}}
\end{eqnarray*}
and then, solving for $\lambda$ in Eq. \ref{lambdacon}, yields the minimizing value of the prices $\bar{p}_i$ given by
\begin{eqnarray}
\label{eqnutility}
m\frac{a_id_i}{\bar{p}_i^{a_i}}=\bar{p}_iq_i\sum_{j=1}^N\frac{a_jd_j}{\bar{p}_j^{a_i}} 
\end{eqnarray}

\subsection{Model Utility Function}
To obtain the utility function,  the value of $\bar{p}_i$ is substituted into the demand function given in Eq. \ref{modeldemand}. To explicitly obtain the minimizing value of the prices $\bar{p}_i$, assume for simplicity that $a_i=a$; then, solving Eq. \ref{eqnutility} yields  $\bar{p}_i$ the following 
\begin{eqnarray}
\label{finalpr}
\bar{p}_i=C \left(\frac{d_i}{q_i} \right)^{1/(a+1)}~~;~~C=\frac{m}{\sum_i {d_i^{1/(a+1)} q_i^{a/(a+1)}}}
\end{eqnarray}
 and yields the utility function 
\begin{eqnarray}
\label{utilitysqrt}
&&\mathcal{U}[\textbf{q}]= \mathcal{D}[\bar{\textbf{p}}(\textbf{q},m)]=\frac{m^{1-a}}{2}\left(\sum_i d_i^{1/(a+1)} q_i^{a/(a+1)}\right)^{a+1}~~;~~a,d_i>0  
\end{eqnarray}
For a single commodity, the utility function is given by
\begin{eqnarray}
\label{utilityone}
&&\mathcal{U}[q]= \mathcal{D}[\bar{p}(q,m)]=\frac{m^{1-a}d}{2}  q^{a} 
\end{eqnarray}

Note that the utility function depends on the budget constraint $m$ for all values of $a$ except the special case of $a=1$. The utility function is sometimes considered to be independent of the budget constraint; however, the utility function being a function of $m$ is also consistent since the budget constraint clearly influences the preferences of the aggregate consumer.

The model utility function given in Eq. \ref{utilitysqrt} clearly fulfills the requirement given in Eq. \ref{fundproputility}, namely that
\begin{eqnarray}
&&\frac{m^{1-a}}{2}\left(\sum_i d_i^{1/(a+1)} q_i^{a/(a+1)}\right)^{a+1} >\frac{m^{1-a}}{2}\left(\sum_i d_i^{1/(a+1)} \tilde{q}_i^{a/(a+1)}\right)^{a+1} \\
&&~~~~~~~~~~~~~\text{If and only if}~~q_i>\tilde{q}_i~~ \text{for any}~i
\end{eqnarray} 

Keeping in mind that $q_i>0$, the marginal utility, as expected, is positive and it can be shown that for the utility function given in Eq. \ref{utilitysqrt}
\begin{eqnarray}
&&\frac{\partial \mathcal{U}[\textbf{q}]}{\partial q_i}>0
\end{eqnarray}
The utility function given in Eq. \ref{utilitysqrt} exhibits diminishing marginal utility since
\begin{eqnarray}
\label{marutility}
&&\frac{\partial^2 \mathcal{U}[\textbf{q}]}{\partial q_i^2}<0
\end{eqnarray} 

\section{The Supply function} \label{sec:supply}
The supply function $\mathcal{S}$ depends on the prices of commodities, and determines the quantity of a commodity that producers are willing and able to sell for a given price. Hence $\mathcal{S}=\mathcal{S}[\textbf{p}]$, where $\textbf{p}=(p_1,p_2,..,p_N)$. Clearly, the supply function must be an increasing function of prices, since the higher the prices, the more is the producer of a commodity willing to supply the said commodity. Hence
\begin{eqnarray}
\label{fundpropsupply}
p'_i>p_i~~~ \text{if and only if}~~~ \mathcal{S}[p_1,p_2,..,p'_i,..p_N]>\mathcal{S}[p_1,p_2,..,p_i,..p_N]
\end{eqnarray}

The supply function $\mathcal{S}[\textbf{p}]$ must be dimensionless since the relative supply of qualitatively different commodities is aggregated into a single supply function. Hence, the prices of commodities need to enter the supply function in dimensionless combinations. 

The supply function is taken to be a function \textit{independent} of the demand function, and is fixed by the drive of capital in seeking returns by engaging in production. This assumption could change in a planned economy and is not explored in this paper. 

The total \textit{supply function} in terms of commodity  quantities $q_i$ is given by
\begin{eqnarray}
\label{consump}
\mathcal{F}[\textbf{q}]=\frac{1}{2}\sum_{i=1}^N \alpha_iq_i  
\end{eqnarray} 
The coefficients $\alpha_i$ are determined by the relative importance of quantity $q_i$ in the supply to the  market of the collection of $N$ commodities.

The total profit from the production of commodities is given by
\begin{eqnarray}
\label{tporfit}
\pi[\textbf{q}]=\sum_{i=1}^N p_iq_i-C[\textbf{q}]
\end{eqnarray} 
$C[\textbf{q}]$ is the cost function. 

The supply function $\mathcal{F}[\textbf{q}]$, namely the quantities produced, is fixed by the company producing only such quantities of commodities that maximizes its profit. More precisely, given the value of $p_i$, the quantity $q_i$  is determined by maximizing $\pi[\textbf{q}]$. Hence
\begin{eqnarray}
\label{maxprofit}
\frac{\partial \pi[\textbf{q}]}{\partial q_i}\Big{|}_{\textbf{q}=\bar{\textbf{q}}}=0~~\Rightarrow~~\bar{\textbf{q}}=\bar{\textbf{q}}(\textbf{p})
\end{eqnarray} 
The supply function in terms of market prices $\textbf{p}$, denoted by $\mathcal{S}[\textbf{p}]$, is given by the following
\begin{eqnarray}
\label{supplqp}
\mathcal{S}[\textbf{p}]=\mathcal{F}[\bar{\textbf{q}}(\textbf{p})]
\end{eqnarray} 

Consider the cost function 
\begin{eqnarray}
\label{modelconsump}
&&C[\textbf{q}]=\sum_{i=1}^N \frac{b_i}{1+b_i} \beta_iq_i^{1+1/b_i}
\end{eqnarray} 
where $\beta_i$ and $b_i$ are related to the cost of producing the commodities, the price of risk in undertaking production plus the expected return on invested capital. The profit of the company is the following
\begin{eqnarray}
\label{modelcost}
&& \pi[\textbf{q}]=\sum_{i=1}^N p_iq_i-\sum_{i=1}^N \frac{b_i}{1+b_i} \beta_iq_i^{1+1/b_i}
\end{eqnarray}
Maximizing profit yields
\begin{eqnarray}
\label{modelmaxprofit}
\frac{\partial \pi[\textbf{q}]}{\partial q_i}\Big{|}_{\textbf{q}=\bar{\textbf{q}}}=0=p_i-\beta_i\bar{q}_i^{1/b_i}~~\Rightarrow ~~\bar{q}_i=\left(\frac{p_i}{\beta_i}\right)^{b_i}
\end{eqnarray}
Hence, from Eqs. \ref{consump}, \ref{supplqp} and \ref{modelmaxprofit}, the supply function is given by
\begin{eqnarray}
\label{maxprofit}
\mathcal{S}[\textbf{p}]=\mathcal{F}[\bar{\textbf{q}}(\textbf{p})]=\frac{1}{2}\sum_{i=1}^N\alpha_i\bar{q}_i =\frac{1}{2}\sum_{i=1}^N \alpha_i\left(\frac{p_i}{\beta_i}\right)^{b_i}
\end{eqnarray} 
Let $\alpha_i/\beta_i^b=ms_i$; the supply function is given by
\begin{eqnarray}
\label{modelsupply}
\mathcal{S}[\textbf{p}]=\frac{m}{2}\sum_{i=1}^N s_ip_i^{b_i} ~~;~~b_i,s_i>0 
\end{eqnarray}
The supply function is \textit{scaled} by the budget constraint $m$ of the aggregate consumer. The scaling is done with the view that the price offered for a commodity is meaningful only if the consumer has non-zero buying power. In the absence of consumer buying power, the effective supply of all commodities is zero.

The supply function is dimensionless and positive valued, that is $\mathcal{S}[\textbf{p}]>0$, with parameter $ms_i$ determining the relative quantity of supply of commodity $i$ with price $p_i$. 

\section{Price versus quantity in standard microeconomics}
To avoid mixing up two different results, for standard microeconomics \cite{varian}  market prices are denoted by $\textbf{p}^*$, the quantity traded is denoted by $\textbf{q}^*$ ; the market price and quantity traded is found by equating the supply of commodities (by the producers) to be equal to the demand for these commodities (by the consumers). In contrast to standard microeconomics, in the statistical microeconomic approach, market prices are denoted by $\textbf{p}_0$, the quantity traded is denoted by $\textbf{q}_0$;  the relation of prices to quantities is not given by equating supply with demand, but instead is given by minimizing the microeconomic potential $\mathcal{V}[\textbf{p}]$ given in Eq. \ref{defpot} -- wand is discussed in Section \ref{micromarket}. 

\begin{figure}[h]
  \centering
  \epsfig{file=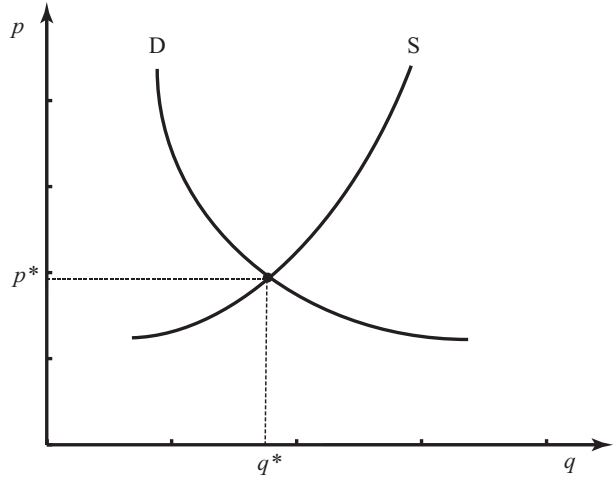, height=7cm, angle=0}
  \caption{The supply and demand for price and quantity, for one commodity. For large $q_i$, The demand price as a function of quantity goes as $p_i\simeq q_i^{-a_i}$ and the supply prices goes as $p_i\simeq q_i^{b_i}$.}
   \label{microsupplydemandquantity}
\end{figure}

For the special case of $a_i=a$, recall from Eq. \ref{finalpr}, the price for given quantity of a commodity that a consumer is willing to pay is the following 
\begin{eqnarray*}
\bar{p}_i=\frac{m}{\sum_j {d_j^{1/(a+1)} q_j^{a/(a+1)}}}\left(\frac{d_i}{q_i} \right)^{1/(a+1)}~~:~~\text{Demand price versus quantity}
\end{eqnarray*}
Furthermore, the price that a supplier is willing to sell a quantity of commodities, from Eq. \ref{modelmaxprofit}, is the following 
\begin{eqnarray*}
p_i=\beta_i\bar{q}_i^{1/b}~~:~~\text{Supply price versus quantity}
\end{eqnarray*}

Figure \ref{microsupplydemandquantity} shows the relation of quantity to price for supply and demand of a single commodity, with the intersection of the two yielding market price and quantity $\textbf{q}^*,\textbf{p}^*$.

\begin{figure}[h]
  \centering
  \epsfig{file=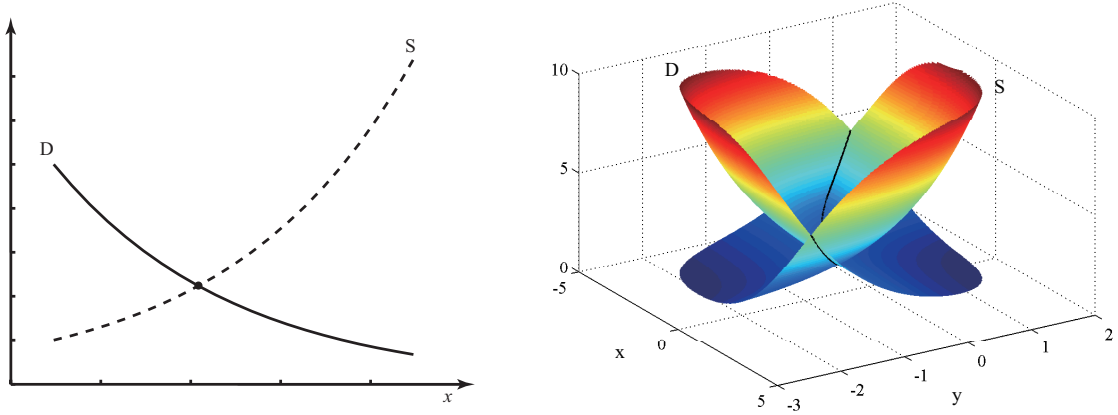, height=6cm, angle=0}
  \caption{(a) Microeconomic supply and demand function for one $p_1=e^x$. b) Supply and demand function for one $p_1=e^x$ and $p_2=e^y$. The unique intersection point of the supply and demand curve is at the minimum of the line of intersection of the supply and demand surfaces.}
   \label{microsupplydemanddiag3}
\end{figure}

Hence, setting the quantity in demand for a commodity $q_i$, at market price $\textbf{p}^*$, to be equal to the supply given by $\bar{q}_i$, yields, for $q_i=q_i^*=\bar{q}_i$ from above  Eqs. \ref{finalpr} and \ref{modelmaxprofit}, the  quantities $q_i^*$ sold in the market are given by
\begin{eqnarray}
\label{pricquant}
&&\frac{m}{\sum_i {d_i^{1/(a+1)} (q^*_{i})^{a/(a+1)}}} \left(\frac{d_i}{q^*_{i}} \right)^{1/(a+1)}=p^*_{i}=\beta_i(q^*_{i})^{1/b} 
\end{eqnarray}
Solving the nonlinear equation given in Eq. \ref{pricquant} yields the following
\begin{eqnarray}
\label{pricquantgen}
 && \textbf{q}^*=\textbf{q}^*(\beta,\textbf{d})~~;~~  \textbf{p}^*=\textbf{p}^*(\beta,\textbf{d})
\end{eqnarray}
Eq. \ref{pricquant}  yields quantities $\textbf{q}^*$  that are bought by the aggregate consumer at prices $\textbf{p}^*$.

A graph of the supply and demand as a function of price,  for one commodity and two commodities is shown in Figure \ref{microsupplydemanddiag3}.

\section{Microeconomic potential} \label{micromarket}
It is \textit{postulated} that the interplay of the supply and demand functions determines the stationary prices of commodities. The trade off between supply and demand is encoded in the microeconomic potential $\mathcal{V}[\textbf{p}]$, given in Eq. \ref{defpot} and defined in terms of the market prices of commodities and as the \textit{sum }of the demand and supply function
\begin{eqnarray*}
\mathcal{V}[\textbf{p}]=\mathcal{D}[\textbf{p}]+\mathcal{S}[\textbf{p}]
\end{eqnarray*}
The potential $\mathcal{V}[\textbf{p}]$ is dimensionless since both $\mathcal{D}[\textbf{p}]$  and $\mathcal{S}[\textbf{p}]$  are dimensionless.

The dependence of the demand and supply function on the market prices of commodities, given in Eqs. \ref{fundpropdemand} and \ref{fundpropsupply} yields the following general limiting behavior for the microeconomic potential
\begin {eqnarray}
\label{microvlimit}
 \mathcal{V}[\textbf{p}] \to \left\{
 \begin{array}{l}
  \mathcal{D}[\textbf{p}]  \to \infty ~~;~~p_i \to 0\\
   \mathcal{S}[\textbf{p}] \to \infty~~;~~p_i \to \infty
\end{array}
\right.   
\end {eqnarray}
The asymptotic behavior given in Eq. \ref{microvlimit} is due to the competing dependence of demand and supply on market prices. Hence,  a \textit{minimum} value  for  $\mathcal{V}[\textbf{p}]$ always exists, for some prices $\textbf{p}_0$ . It is shown in Eq. \ref{microcoavgprconstant} that $\textbf{p}_0$, to leading order for the model chosen, is equal to the average value of market prices. The value of the minimizing price vector $\textbf{p}_0$  is given by the following
\begin{eqnarray}
\label{micropotmin}
\frac{\partial \mathcal{V}[\textbf{p}]}{\partial p_i}\Big{|}_{\textbf{p}=\textbf{p}_0}=0\\ \nonumber\\
\label{micropotminsupdem}
\Rightarrow \frac{\partial \mathcal{D}[\textbf{p}]}{\partial p_i}\Big{|}_{\textbf{p}=\textbf{p}_0}=-\frac{\partial \mathcal{S}[\textbf{p}]}{\partial p_i}\Big{|}_{\textbf{p}=\textbf{p}_0}
\end{eqnarray}
In other words, as can be seen from Eq. \ref{micropotminsupdem}, a minimum value of the potential $\mathcal{V}[\textbf{p}]$ is attained at price vector $\textbf{p}_0$ when a small variation of prices yields a change of demand that is exactly the opposite to  change of supply.

\begin{figure}[h]
  \centering
  \epsfig{file=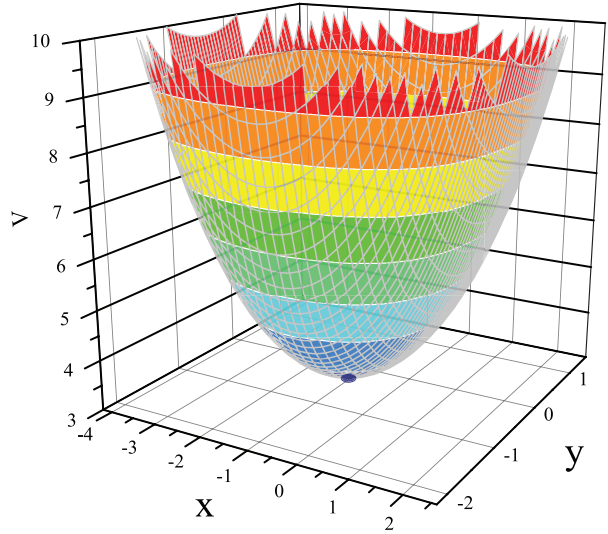, height=7cm, angle=0}
  \caption{Microeconomic potential $\mathcal{V}[\textbf{p}]$ for two prices $p_1=e^x$ and $p_2=e^y$ showing a unique minimum value, given by the dot at the minimum of the surface.}
  \label{micrpotdiag}
\end{figure}

As can be seen from Figure \ref{micrpotdiag} the microeconomic potential has a surface that is required for the minimization that yields market prices $\textbf{p}_0$; in contrast, for the standard microeconomic, only the intersection of the supply and demand curve are relevant, as shown in  Figure \ref{microsupplydemanddiag3}(b). 

Whether the minimizing prices $\textbf{p}_0$ are unique or not depends on the model chosen for $\mathcal{V}[\textbf{p}]$; since market prices are known to be unique a requirement for all models in microeconomics is that they yield a unique value for $\textbf{p}_0$.

Note that the price vector $\textbf{p}_0$ that minimizes the potential $\mathcal{V}[\textbf{p}]$ has no relation to the minimization carried out in Section \ref{sec:demandutily} since in going from the demand to the utility function one is maximizing the demand function that is constrained by the budget $m$. In contrast, the minimization of  $\mathcal{V}[\textbf{p}]$ is \textit{unconstrained} and fixes the market price as a function of the parameters of the model potential.

One would like to have a time independent potential since then one can make unique predictions of future movement of market prices of commodities. One can also introduce explicit time dependence in the potential  $\mathcal{V}[\textbf{p}]$ to reflect major scheduled announcements such of quarterly industrial output, employment figures, yearly budgets and so on. 

\section{Model of microeconomic potential}
For the model chosen for the demand and supply functions given in Eqs. \ref{modeldemand} and \ref{modelsupply} respectively, the microeconomic potential is given by
\begin{eqnarray}
\label{micropotenmodel}
\mathcal{V}[\textbf{p}]&=&\mathcal{D}[\textbf{p}]+\mathcal{S}[\textbf{p}]\nonumber\\
   &=&\frac{m}{2}\left[\sum_{i=1}^N\frac{d_i}{p_i^{a_i}} + \sum_{i=1}^N s_ip_i^{b_i}\right]~;~~d_i,s_i>0 ~;~~a,b>0 
\end{eqnarray}
The model microeconomic potential has the expected asymptotic behavior Eq. \ref{microvlimit}, and is realized in the following manner for the model chosen
\begin {eqnarray}
 \mathcal{V}[\textbf{p}] \to \left\{
 \begin{array}{l}
  \mathcal{D}[\textbf{p}] \simeq 1/p_i^{a_i} \to \infty ~~;~~p_i \to 0\\
   \mathcal{S}[\textbf{p}] \simeq p_i^{b_i} \to \infty~~;~~p_i \to \infty
\end{array}
\right.   
\end {eqnarray}

Figure \ref{micrpotdiag} shows the shape of $\mathcal{V}[\textbf{p}]$ for the model given in Eq. \ref{micropotenmodel}; note the important feature of  $\mathcal{V}[\textbf{p}]$ that it has a (unique) global minimum at $\textbf{p}_0$. The value of $\textbf{p}_0$ is obtained by minimizing $\mathcal{V}[\textbf{p}]$ and, from Eqs. \ref{micropotenmodel} and \ref{micropotmin}, yields the following
\begin{eqnarray}
&&\frac{\partial \mathcal{V}[\textbf{p}]}{\partial p_i}\Big{|}_{\textbf{p}=\textbf{p}_0}=0  ~~\Rightarrow~~  -a_i\frac{d_i}{p_{0i}^{a_i+1}} + b_i s_ip_{0i}^{b_i-1}=0\nonumber\\
\label{minimamicropotenmodel}
&&p_{0i}=\left(\frac{a_id_i}{b_is_i}\right)^{1/(a_i+b_i)}
\end{eqnarray}

In standard microeconomic theory, the market prices $\textbf{p}^*$ are fixed by equating demand to supply, shown graphically in Figure \ref{microsupplydemanddiag3}; for the model being considered, this yields the following
\begin{eqnarray}
\label{demandsuppleq}
&&\mathcal{D}[\textbf{p}^*]=\mathcal{S}[\textbf{p}^*] ~~\Rightarrow~~\frac{d_i}{(p^*_{i})^{a_i}} =  s_i(p^*_{i})^{b_i}~~\Rightarrow~~ p^*_{i}=\left(\frac{d_i}{s_i}\right)^{1/(a_i+b_i)}
\end{eqnarray}

Equating the supply and demand functions, shown graphically in Figure \ref{microsupplydemanddiag3}, yields the average price $\textbf{p}^*$ different from the result of $\textbf{p}_0$ given by minimizing the potential $\mathcal{V}[\textbf{p}]$ as given in  Eq. \ref{minimamicropotenmodel}. It is only  for the very special of a single commodity and with potential of $a=1=b$ that the two approaches yield the same answer.

Note that the expressions obtained for $\textbf{p}_0$ and  $\textbf{p}^*$ are very different than the relation of market price and quantity obtained in Eq. \ref{pricquantgen}: in Eqs. \ref{minimamicropotenmodel} and \ref{demandsuppleq} the market price is obtained in terms of the parameters of the supply and demand function whereas Eq. \ref{pricquantgen} yields the locus $\textbf{p}^*(\textbf{q}^*)$ -- namely, the market price of commodities and the quantity of these commodities sold on the market.

\subsection{Potential versus supply and demand}
The concept of a potential carries more information about prices than the intersection of the supply and demand curve. The following are some of the reasons.
\begin{itemize}
\item Figure \ref{microsupplydemanddiag3}(b) shows two surfaces, namely that of the supply and demand surfaces  whereas, in contrast Figure \ref{micrpotdiag} shows only a single surface. Both figures can, in principle,  be used for determining the stationary prices. But,  as can be seen by comparing Eqs. \ref{minimamicropotenmodel} and \ref{demandsuppleq}, these prices are quite different. 
\item Equating the supply and demand functions considered separately yields only a single `market' price with no information about what are the possible variations about the market price. In contrast, being the minimum of the potential, the potential also contains information about the commodity prices  in the \textit{neighborhood} of market prices  $\textbf{p}_0$ as well as commodity prices that are far from the market  prices.
\item The statistical variation of market prices needs information on how the demand and supply compete to set prices far from equilibrium since all values of prices are allowed in computing the expected value of market observed prices. The potential, by combining supply and demand into one function, models the competing influences that supply and demand have on commodity prices.
\item Together with the kinetic term for prices, discussed in next Section, the potential plays a central role in determining the statistical evolution of commodity prices near -- as well far from -- its average value.
\end{itemize}

\section{Microeconomic kinetic term}
The dynamics of market prices is encoded in the kinetic component $\mathcal{T}[\textbf{p}]$ of the Lagrangian given in Eq. \ref{defpot}.

Market prices undergo a dynamical evolution and hence depend on time, namely $p_i=p_i(t)$. Furthermore, similar to the microeconomic potential, the kinetic component is \textit{assumed} to depend linearly on budget $m$ -- since one expects no dynamics for a market in which the consumer has no buying power. The linear dependence of $\mathcal{T}[\textbf{p}]$ on the budget $m$ is taken for simplicity and can be generalized. 

A detailed empirical study of both interest rates \cite{bebcup2} and of equity prices \cite{bebcyeqt} shows that the Lagrangian depends on both, the \textit{velocity} and \textit{acceleration} of the underlying security. Since the markets for commodities are connected to the financial and capital markets, one expects that the dynamics of commodities should have the same behavior as interest rates and equities.

Since prices are always positive, consider the exponential parametrization
\begin{eqnarray}
\label{defexpvar}
p_i(t)=p_{0}e^{x_i(t)}~~;~~-\infty \le  x_i \le +\infty
\end{eqnarray}
where $p_0$ is a constant quantity with the dimension of \$.

The kinetic term for the market prices of commodities, in tandem with the capital and debt markets,  is  taken to be the following
\begin{eqnarray}
\label{microkinetic}
&&\mathcal{T}[\textbf{p}] =\frac{m}{2}\sum_{i,j=1}^N \left[L_{ij}\frac{\partial^2 x_i}{\partial t^2}\frac{\partial^2 x_j}{\partial t^2}+\tilde{L}_{ij}\frac{\partial x_i}{\partial t}\frac{\partial x_j}{\partial t}\right]
\end{eqnarray}
The form of the time dependence given in Eq. \ref{microkinetic} yields many features for the dynamics of microeconomics that are are not present in quantum and classical mechanics, for which the Lagrangian depends on only the particle's velocity. 

The kinetic term for market prices has the following cardinal properties.
\begin{itemize}
\item The kinetic term $\mathcal{T}[\textbf{p}]$ does not depend on   $p_0$, but rather, depends only on the relative instantaneous changes in the price vector.
\item The second order derivative term in the kinetic term requires \textit{four} boundary conditions to specify a classical solution of the Lagrangian \cite{cythesis}. 
\item The market prices undergo a correlated time evolution that is determined by the matrix of parameters given by $L_{ij}$ and $\tilde{L}_{ij}$. 
\item The budget constraint influences the correlation on market prices that in turn determines the uptake of commodities $q_i$ by the aggregate consumer.
\end{itemize}

\section{Microeconomic Feynman Path Integral}
The Lagrangian of a system  determines the evolution of a dynamical system and for market prices represents all the factors determining its evolution. In particular, the interplay and competition of demand, supply with the `kinetic energy' of market prices is encoded in the Lagrangian. 

The Lagrangian, from Eq. \ref{microlagrang}, is given by the sum of the kinetic and potential factors and yields
\begin{eqnarray*}
&&\mathcal{L}(t) = \mathcal{T}[\textbf{p}(t)]+\mathcal{V}[\textbf{p}(t)]
\end{eqnarray*}

The action functional determines the dynamics (time evolution) of market prices and, from Eq. \ref{microaction} is  given by
\begin{eqnarray*}
\mathcal{A}[\textbf{p}]=\int_{-\infty}^{+\infty} dt\mathcal{L}(t)=\int_{-\infty}^{+\infty} dt \Big(\mathcal{T}[\textbf{p}(t)]+\mathcal{V}[\textbf{p}(t)]\Big)
\end{eqnarray*}
The model chosen for the potential and kinetic parts of the Lagrangian yields, from Eqs. \ref{micropotenmodel} and \ref{microkinetic} the following
\begin{eqnarray}
\label{microlagrangmodel}
\mathcal{L}(t) = \frac{m}{2}\sum_{i,j=1}^N \left[L_{ij}\frac{\partial^2 x_i}{\partial t^2}\frac{\partial^2 x_j}{\partial t^2}+\tilde{L}_{ij}\frac{\partial x_i}{\partial t}\frac{\partial x_j}{\partial t}\right]+\frac{m}{2}\sum_{i=1}^N\frac{d_i}{p_i^{a_i}} + \frac{m}{2}\sum_{i=1}^N s_ip_i^{b_i}~~
\end{eqnarray}
The  Lagrangian  given in Eq. \ref{microlagrangmodel}  is nonlinear, since prices (and quantities) are always positive -- and hence represented by exponential variables as in Eq. \ref{defexpvar}. 

For the case of a single commodity, let the price be $p=p_0e^x$; the Lagrangian given in Eq. \ref{microlagrangmodel} reduces to the following
\begin{eqnarray}
\label{microlagrangmodelonedoff}
\mathcal{L}(t) = \frac{m}{2}\left[L\left(\frac{\partial^2 x}{\partial t^2}\right)^2+\tilde{L}\left(\frac{\partial x}{\partial t}\right)^2\right]+\frac{m}{2}\left[\frac{d}{p_0}e^{-ax} +  sp_0e^{bx}\right]~~;~~p=p_0e^x>0
\end{eqnarray}

The stochastic processes driving the market prices are modeled in analogy with statistical mechanics, for which the  particles' deterministic positions and velocities are generalized to random positions and velocities. Similarly, in statistical microeconomics, it is postulated that all prices are random variables; the \textbf{joint probability distribution} for the market prices to have a particular evolution $\{\textbf{p}(t): -\infty \le t \le +\infty\}$, given in Eq. \ref{boltzprop}, has the following properly normalized form 
\begin{eqnarray}
\label{micropdf}
&&\frac{e^{-\mathcal{A}[\textbf{p}]}}{Z}
\end{eqnarray}
Note that the statistical weight provided by $\exp\{-\mathcal{A}[\textbf{p}]\}/Z$ determines which random histories of prices are important and which are not.

The normalization $Z$ is given by the Feynman path integral
\begin{eqnarray}
\label{micropartion}
&&Z=\prod_{i=1}^{N}\prod_{t=-\infty}^{+\infty}\int_{-\infty}^{+\infty}\frac{dp_i(t)}{p_i(t)}e^{-\mathcal{A}[\textbf{p}]}\equiv \int \frac{Dp}{p} ~e^{-\mathcal{A}[\textbf{p}]}~~:~~\text{Feynman path integral}
\end{eqnarray}
The correlation function of market prices is given by the expectation value of the product of prices, computed by summing over all possible histories of market prices using the path integral, and is given by the following
\begin{eqnarray}
\label{microcorrefns}
&&E[p_{i_1}(t_1)p_{i_2}(t_2)..p_{i_N}(t_N)]=\frac{1}{Z}\int \frac{Dp}{p} e^{-\mathcal{A}[\textbf{p}]}p_{i_1}(t_1)p_{i_2}(t_2)..p_{i_N}(t_N)
\end{eqnarray}

The path integral, for exponential variables $x_i$, defined in Eq. \ref{defexpvar},  is given by
\begin{eqnarray}
\label{micropartionxx}
&&Z=\prod_{i=1}^{N}\prod_{t=-\infty}^{+\infty}\int_{-\infty}^{+\infty}dx_i(t)e^{-\mathcal{A}[p_{0},x_i]} \equiv \int DX e^{-\mathcal{A}[p_{0i}e^{x_i}]}
\end{eqnarray}
with the correlations given by
\begin{eqnarray*}
&&E[p_i(t_1)p(t_2)..p(t_N)]=\frac{p_0^N}{Z}\int DX e^{-\mathcal{A}[p_{0},x_i]}e^{x_{i_1}(t_1)}e^{x_{i_2}(t_2)}..e^{x_{i_N}(t_N)}
\end{eqnarray*}
Note the expression for $Z$ given in Eq. \ref{micropartionxx} is \textit{exact} and no approximation has been made; rather  a change of variables has been made from $p_i$ to $x_i$, where the new set of variables are more suitable for the perturbative study of the microeconomic path integral.

The path integral given in Eq. \ref{microcorrefns} is nonlinear and nontrivial. The path integral can be studied numerically using Monte Carlo and other well known methods. In many cases, the numerical approach is necessary for studying non-perturbative features that are inaccessible to other methods.  

\section{Perturbation expansion}
Analytic computations of the path integral is one of the standard methods for understanding the main qualitative features of a system represented by a path integral. Given that the path integral for prices is nonlinear, an exact solution is close to impossible; the best that one can do is to develop an approximation scheme, with the standard approach being a perturbation expansion in some small parameter -- with higher and higher terms in the expansion parameter being more and more accurate.

In the action functional $\mathcal{A}[\textbf{p}]$,  all the parameters were scaled so that the inverse of the total budget, namely $1/m$, provides a small expansion parameter for the following reason. For the case of $m>>1$, the path integral given in Eq. \ref{microcorrefns} is dominated by the values of $p(t)$ for which the integrand $\exp\{-\mathcal{A}[\textbf{p}]\}$ is a \textit{maximum}, or equivalently, for which $\mathcal{A}[\textbf{p}]$ is a minimum. In particular, the path integral can be expanded as a powers series in $1/m$ and can generate an expansion in powers of $1/m$   for all quantities of interest. The inverse of the budget $1/m$ behaves like Planck's constant of quantum mechanics, and a power series expansion in terms of $1/m$ is called a semi-classical expansion. 

The path (historical evolution) of $p(t)$ that minimizes $\mathcal{A}[\textbf{p}]$ is given by\footnote{In the context of quantum mechanics, the path of the prices that minimizes the action, namely $p_c(t)$ is called the classical solution.}
\begin{eqnarray}
\label{classical}
&&\frac{\delta \mathcal{A}[\textbf{p}_c]}{\delta p_i(t)}\equiv \frac{\delta \mathcal{A}[\textbf{p}]}{\delta p_i(t)}\Big{|}_{(\textbf{p}(t)=\textbf{p}_c(t))}=0
\end{eqnarray} 

Non-linear actions, such as the one for prices given in Eq. \ref{microlagrangmodel}, can have a minimum value for solutions $\textbf{p}_c(t)$ that have non-trivial time dependence, and are called kinks or instantons. Kinks are time dependent solutions of Eq. \ref{classical} that connect nontrivial initial and final boundary conditions.

For simplicity, consider the case where the  prices $\textbf{p}_c(t)$ that minimize the action  are \textit{time independent} (constant). For time independent $\textbf{p}_c$, the action functional $\mathcal{A}[\textbf{p}_c]$ is equal to the time integral of the potential $\mathcal{V}[\textbf{p}_c]$ and the minimum of the action functional is given by the minimum of the potential $\mathcal{V}[\textbf{p}_c]$. 

\begin{figure}[h]
  \centering
  \epsfig{file=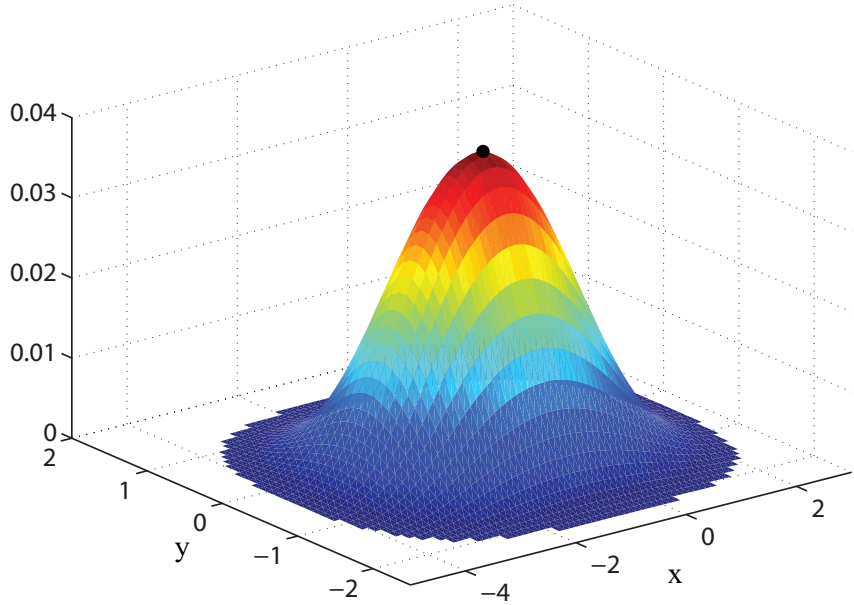, height=10cm, angle=0}
  \caption{The function $\exp\{-\mathcal{V}[\textbf{p}]\}$ near the minimum value of the  microeconomic potential.}
  \label{micropotmax}
\end{figure}

As shown in Figure \ref{micrpotdiag}, the microeconomic potential  $\mathcal{V}[\textbf{p}]$ has a \textit{minimum} at $\textbf{p}_c(t)=\textbf{p}_0$; hence , $\exp\{-\mathcal{V}[\textbf{p}]\}$ has a \textit{maximum} for the value of market prices  $\textbf{p}_c(t)=\textbf{p}_0$, as shown in Figure \ref{micropotmax}. Hence, similar to Figure \ref{micropotmax}, $\exp\{-\mathcal{A}[\textbf{p}]\}$ --  the integrand of the path integral -- has a maximum for which the potential has a minimum, namely at $\textbf{p}_c(t)=\textbf{p}_0$. 

A perturbative expansion of the path integral is based on the statistical fluctuations of the price vector in the neighborhood of the minimum value of the microeconomic potential, namely in the neighborhood of $\textbf{p}=\textbf{p}_0$. In terms of the exponential variables given in Eq. \ref{defexpvar}, the minima of the microeconomic potential is given by the value of $\bar{x}_i$; the minima $\textbf{p}_0$ given by Eq. \ref{minimamicropotenmodel}  yields the following
\begin{eqnarray}
\label{pox}
&&p_i=p_{0}e^{x_i}~~;~~p_{0i}=p_0e^{\bar{x}_i}=\left(\frac{a_id_i}{b_is_i}\right)^{1/(a_i+b_i)}
\end{eqnarray} 

By expanding the path integral about the value of $\bar{x}_i$ as given in Eq. \ref{pox}, the path integral in Eq. \ref{micropartionxx}  yields a perturbation expansion in terms of $1/m$ for the following reason. Under normal market conditions it is expected that prices have \textit{small fluctuations} about the minimum value of the action $\mathcal{A}[\bar{x}_i]$. In fact, it is shown later in Eq. \ref{avgx2} that, near the maximum of the action given by $\mathcal{A}[\bar{x}_i]$, the magnitude of the integrations variables is given by
\begin{eqnarray}
\label{flucm}
x_i=\bar{x}_i+O(\sqrt{\frac{1}{m}})
\end{eqnarray} 
The result given in Eq. \ref{flucm} is intuitively the following: for a consumer with a large budget $m$, prices are very near $p_0\exp\{\bar{x}_i\}$  and fluctuate very little since the large budget allows the consumer to buy any commodity he or she wishes. However, as the budget becomes smaller and smaller, the fluctuations in the prices become larger and larger since the consumer now has to make a choice, buying some commodities and foregoing others; this leads to large changes in the uptake of different commodities and hence introducing large random variations in the prices.

Making a functional change of variables from $x_i(t)$ to $y_i(t)$ 
\begin{eqnarray}
\label{chngxtoy}
x_i(t)=\bar{x}_i+y_i(t)
\end{eqnarray}
The path integral measure is invariant under a shift and hence $DX=DY$; hence,  Eq. \ref{micropartionxx} yields the path integral
\begin{eqnarray}
\label{zpathy}
&&Z=\int DY e^{-\mathcal{A}[\bar{x}_i+y_i(t)]}
\end{eqnarray}
The path integral given in Eq. \ref{zpathy} allows for an expansion of the action $\mathcal{A}[\bar{x}_i+y_i(t)]$ in a Taylor power series of $y_i$ about the maxima of $\mathcal{A}[\bar{x}_i]$ at $\bar{x}_i$ and will be shown to yield a convergent expansion of all the correlation functions in powers of $1/m$. 

In the expansion of $\mathcal{A}[\bar{x}_i+y_i(t)]$ about its minimum value $\mathcal{A}[\bar{x}_i]$, there are no terms that are linear in $y_i$ due to Eq. \ref{classical}; the first term is a constant and the next leading term is quadratic in $y_i$, and with the remaining terms in the expansion of the action all having powers that are $y_i^3$ and higher. Hence, the action functional has the following expansion
\begin{eqnarray}
\label{actionexpans}
&&\mathcal{A}[p_0; x_i]=\mathcal{A}_1[\bar{x}]+\mathcal{A}_2[\bar{x};y^2]+\mathcal{A}_I[\bar{x};y^3]
\end{eqnarray}
with
\begin{eqnarray*}
&&\mathcal{A}_1[\bar{x}]:~~~~~\text{Constant independent of}~y_i\\
&&\mathcal{A}_2[\bar{x};y^2]:~\text{Quadractic function of}~y_i\\
&&\mathcal{A}_I[\bar{x};y^3]:~\text{Cubic and higher order function of}~y_i
\end{eqnarray*}

The integration variables $y_i(t)$ are of $O(\sqrt{\frac{1}{m}})$ and hence the successive terms in the expansion of the action $\mathcal{A}[p_0; x_i]$ in Eq. \ref{actionexpans} are of smaller and smaller magnitude.

Note the expansion of the action about $\mathcal{A}[\bar{x}]$ is valid \textit{only} for $m>>1$; for $m\le 1$, the perturbative approach is invalid since there is no longer any sharp and well localized domain of the path integral that gives the dominant contribution. 

If the budget becomes small, such that $m\simeq O(1)$, the statistical fluctuations in the prices $\textbf{y}(t)$ become large and the  perturbation expansion becomes invalid. Of course, the path integral given in Eq. \ref{micropartion} is well defined and convergent for all $m>0$. For the case when $m\simeq 1$, the path integral has to be studied using non-perturbative techniques, and which includes numerically evaluating the path integral.

To illustrate the expansion of the path integral, consider the partition function given in Eq. \ref{micropartionxx}. Expanding the action functional as given above yields the following expansion for the partition function
\begin{eqnarray}
&&Z=\int \frac{Dp}{p}e^{-\mathcal{A}[\textbf{p}]}=e^{-\mathcal{A}_1[\bar{x}]}\int DYe^{-\mathcal{A}_2[\bar{x};y^2]-\mathcal{A}_I[\bar{x};y^3]}\nonumber\\
&&~~~=e^{-\mathcal{A}_1[\bar{x}]}\int DYe^{-\mathcal{A}_2[\bar{x};y^2]}\left[1-\mathcal{A}_I[\bar{x};y^3]+\frac{1}{2!}\mathcal{A}_I^2[\bar{x};y^3]+...\right]\nonumber\\
\label{expanpartion}
&&~~~=z_0+\frac{1}{m}z_1+\frac{1}{m^2}z_1+...
\end{eqnarray}

\section{Expansion of microeconomic potential}
Note that the minimum of the potential chosen in Eq. \ref{micropotenmodel} fixes the price of all $N$ commodities. Writing the potential in terms of variables defined in Eq. \ref{pox} that are appropriate for studying the action functional near its maximum 
\begin{eqnarray}
\label{poxy}
&&p_i=p_{0}e^{\bar{x}_i+y_i}
\end{eqnarray}
yields the following
\begin{eqnarray}
\label{micropotenmodelmin}
\mathcal{V}[\textbf{p}] &=&\frac{m}{2}\sum_{i=1}^N\left[\frac{d_i}{p_i^{a_i}} +  s_ip_i^{b_i}\right]~;~~a_i,b_i,d_i,s_i>0\nonumber \\
\mathcal{V}[\bar{x}; y]&=& \frac{m}{2}\sum_{i=1}^N \left[\frac{d_i}{p_0^{a_i}}e^{-a_i\bar{x}_i}e^{-a_iy_i}+s_ip_0^{b_i}e^{b_i\bar{x}_i}e^{b_iy_i}\right]
\end{eqnarray}

Note Eq. \ref{micropotenmodelmin} is an exact expression for the potential $\mathcal{V}[\textbf{p}]$. Expanding the potential given in Eq. \ref{micropotenmodelmin} as a power series in $y_i$  yields the following
\begin{eqnarray}
\label{micropotenquadra}
 &&   \mathcal{V}[\bar{x}; y] \simeq \mathcal{V}_0+\frac{m}{2}\sum_{i=1}^N \gamma_i x_i^2+O(x^3)\\
 && \gamma_i=\frac{1}{2}\left[\frac{a_i^2d_i}{p_0^{a_i}}e^{-a_i\bar{x}_i}+b^2_is_ip_0^{b_i}e^{b_i\bar{x}_i}\right]\nonumber
\end{eqnarray}
with the constant value of the potential given by
\begin{eqnarray}
 &&\mathcal{V}_0= \frac{m}{2}\sum_{i=1}^N \left[\frac{d_i}{p_0^{a_i}}e^{-a_i\bar{x}_i}+s_ip_0^{b_i}e^{b_i\bar{x}_i}\right]\nonumber
\end{eqnarray}
Note that there is no linear dependence on $y_i$ in the expansion in Eq. \ref{micropotenquadra} since the condition of minimum, given in Eq. \ref{minimamicropotenmodel}, ensures this to be the case.

Since $\bar{x}$ is a constant,  Eqs. \ref{actionexpans} and \ref{micropotenquadra} yield
\begin{eqnarray*}
\mathcal{A}_1[\bar{x}]=&=&\mathcal{V}_0\int_{+\infty}^{+\infty}dt=\text{constant independent of~}y
\end{eqnarray*}

\section{Model of the kinetic term}
For simplicity and tractability, a special choice of the coupling of the first and second order time derivatives is made that, in matrix notation, is given by
\begin{eqnarray}
\label{couplingmatrix}
&&L=D^T\text{diag}(\alpha_1,\alpha_2, ..,\alpha_N) D~~;~~\tilde{L}=D^T\text{diag}(\beta_1,\beta_2, ..,\beta_N) D~~;~~ DD^T=\mathcal{I}~~~~~~
\end{eqnarray}

The kinetic piece of the action functional in Eq. \ref{microaction} is given by the time integral of  $\mathcal{T}[\textbf{p}]$, namely
\begin{eqnarray}
\label{microkineticaction}
&&\int_{-\infty}^{-\infty}dt \mathcal{T}[\textbf{p}(t)]
\end{eqnarray}

To express the kinetic piece in terms of the $x=\bar{x}+y$, we perform an integration by parts for the derivative terms in $\mathcal{T}[\textbf{p}]$ in Eq. \ref{microkineticaction}, and setting all the boundary terms to zero, obtain the following
\begin{eqnarray}
\label{microlagrangexct}
&&~~~~~~~~~~~~~~~~~~~~~~~\int_{-\infty}^{-\infty}dt \mathcal{T}[\textbf{p}(t)]\nonumber\\
&&=\frac{m}{2}\sum_{i,j,k=1}^N \int_{-\infty}^{-\infty}dty_i(t)D^T_{ij}\Big(\alpha_j\frac{\partial^4}{\partial t^4}-\beta_j\frac{\partial^2 }{\partial t^2}\Big)D_{jk}y_k(t)
\end{eqnarray}
Note Eqs. \ref{couplingmatrix} has been used to obtain Eq. \ref{microlagrangexct}.

\section{Gaussian path integration: propagator}\label{sec:pertexpan}
As illustrated in Eq. \ref{expanpartion}, $1/m$ provides a small expansion parameter for the path integral. It should be noted that the exact path integral given in Eq. \ref{micropartionxx} has a Gaussian expansion only for $m<<1$; the reason being that only in this case is the quadratic piece of the action functional $\mathcal{A}_2[\bar{x};y^2]$ the dominant part in the full expansion of the action functional as a power series in $y_i$.

This Section evaluates the propagator for the prices, and which is a central ingredient of the $1/m$ expansion.
 
Expanding the action around the minimum to terms of order $y^2$ yields, from Eqs. \ref{actionexpans} and \ref{micropotenquadra} and \ref{microkinetic}, the following
\begin{eqnarray}
\label{microlagrang5}
&&\mathcal{A}_2[\textbf{p}_{0};x^2] =\frac{m}{2}\sum_{i,j,k=1}^N \int_{-\infty}^{-\infty}dty_i(t)\left[D^T_{ij}\Big(\alpha_j\frac{\partial^4}{\partial t^4}-\beta_j\frac{\partial^2 }{\partial t^2}\Big)D_{jk}\right]y_k(t)\nonumber\\
&&~~~~~~~~~~~~~~~~~~~~~~~~~~~~~~~~~~~+\frac{m}{2}\sum_{i=1}^N \int_{-\infty}^{-\infty}dt\gamma_iy_i^2(t)\nonumber\\
&&\Rightarrow \mathcal{A}_2[\textbf{p}_{0};y^2] \equiv \frac{1}{2}\sum_{i,k=1}^N \int_{-\infty}^{-\infty}dty_i(t)G^{-1}_{ik}(t,t')y_k(t)
\end{eqnarray}
where the inverse of the propagator is given by
\begin{eqnarray}
\label{definvprop}
&&~~~G^{-1}_{ik}(t,t')=m\sum_{j=1}^N \left[D^T_{ij}\Big(\alpha_j\frac{\partial^4}{\partial t^4}-\beta_j\frac{\partial^2 }{\partial t^2}\Big)D_{jk}+ \delta_{j-k}\delta_{i-k}\gamma_k\right]\delta(t-t')~
\end{eqnarray}
The propagator $G_{ij}(t,t')$ is given by
\begin{eqnarray}
\label{propdef}
&&~~~\int_{-\infty}^{-\infty}dt \sum_{k=1}^N G^{-1}_{ik}(t,\xi)G_{kj}(\xi,t')=\delta_{i-j}\delta(t-t')~\\
&& \Rightarrow~~G_{ij}(t,t')\simeq \frac{1}{m}
\end{eqnarray}

The propagator $G_{ij}(t,t')$ has been explicitly worked out in \cite{bebcup2,cythesis} and applied to the study of equity prices in \cite{bebcyeqt}.

For the quadratic action given in Eq. \ref{microlagrang5}, Gaussian path integration yields the following generating functional
\begin{eqnarray}
\label{gausspint}
Z[h]&=&\frac{1}{Z}\int DY \exp\left\{-\mathcal{A}_2[\textbf{p}_{0};y^2]+\sum_i\int dt h_i(t)y_i(t)\right\} \nonumber\\
    &=&\frac{1}{Z}\int DY \exp\left\{- \frac{1}{2}\sum_{i,k=1}^N \int_{-\infty}^{-\infty}dty_i(t)G^{-1}_{ik}(t,t')y_k(t)+\sum_i\int dt h_i(t)y_i(t)\right\} \nonumber\\
   &=&\exp\left\{\frac{1}{2}\sum_{ij}\int dtdt' h_i(t)G_{ij}(t,t')h_j(t')\right\}
\end{eqnarray}

To $O(y^2)$, the expectation value of market prices is given by the following 
\begin{eqnarray}
\label{microcoavgpr}
E[p_i(t)]&=&E[p_{0}e^{x_i(t)}]=\frac{p_{0}}{Z}\int DY e^{-\mathcal{A}[\bar{x}+y(t)]}e^{\bar{x}_i+y_i(t)} \nonumber \\
         &\simeq&\frac{p_{0}e^{\bar{x}_i}}{Z}\int Dx e^{-\mathcal{A}_2[\bar{x};y^2]}p_{0i}e^{y_i(t)} 
\end{eqnarray}
Since $ p_{0i}=p_{0}e^{\bar{x}_i}$, using the rules of Gaussian path integration given in Eq. \ref{gausspint} yields, from Eq. \ref{microcoavgpr}, the following
\begin{eqnarray}
\label{microcoavgpr5}
E[p_i(t)]&\simeq& p_{0i}e^{G_{ii}(t,t)}
\end{eqnarray}

All the parameters of the model are constants and hence
\begin{eqnarray}
\label{proptimeinv}
G_{ij}(t,t')=G_{ij}(t-t')
\end{eqnarray}
From Eqs. \ref{microcoavgpr5} and \ref{proptimeinv}, the average price of the $i$th commodity is hence given by
\begin{eqnarray}
\label{microcoavgprconstant}
E[p_i(t)]&\simeq&p_{0i}e^{G_{ii}(0)}=p_{0i}+O(\frac{1}{m}):~~\text{constant}
\end{eqnarray}
Note, as stated earlier in Eq. \ref{micropotmin}, $p_{0i}$ -- to leading order in $1/m$, -- is the average value of the market price and is a constant.

To $O(1/m^2)$, the \textit{propagator} (correlation function of prices at two different times), using the rules of Gaussian path integration  given in Eq. \ref{gausspint}, yields the following
\begin{eqnarray}
\label{microcorreprop}
E[\ln(\frac{p_i(t)}{p_{0i}})\ln(\frac{p_j(t')}{p_{0j}})]&=&\frac{1}{Z}\int DY e^{-\mathcal{A}[\bar{x}+y]}y_i(t)y_j(t') \nonumber\\
&\simeq&\frac{1}{Z}\int DY e^{-\mathcal{A}_2[\bar{x};y^2]}y_i(t)y_j(t') \nonumber\\
                          &=&G_{ij}(t,t')+O(1/m^2)
\end{eqnarray}
where $G_{ij}(t,t')$ is given in Eq. \ref{propdef}. 

Eq. \ref{microcorreprop} yields the following special case
\begin{eqnarray}
E[y_i(t)^2]&=&G_{ii}(0)=O(\frac{1}{m})\nonumber
\end{eqnarray} 
Hence, the average range of the integration variables where the integrand $e^{-\mathcal{A}_2[\bar{x};y^2]}$ has significant values is given by
\begin{eqnarray}
\label{avgx2}
y_i&=&O(\sqrt{E[y_i(t)^2]})=O(\sqrt{\frac{1}{m}})
\end{eqnarray} 

Eq. \ref{avgx2} shows that the average magnitude of the fluctuations of $y_i^2$ is $O(1/m)$. This is the reason that a perturbation expansion can be generated for all the correlation functions of $p_i$ in increasing powers of $O(1/m)$ (by expanding, in a power series, all the terms in the action functional $\mathcal{A}$ that are of $O(y^3)$ and higher) and leads to an expansion of the path integral as in given Eq. \ref{expanpartion}.

The rules of Gaussian path integration,  using the technique of Feynman diagrams, yield a perturbation expansion for all the correlation functions of commodity prices. In particular, the terms $z_0,z_1, ..$ in the expansion for the partition function $Z$ given in Eq. \ref{expanpartion} can all be evaluated using the rules of Gaussian path integration.

\section{Model calibration and testing}
Every observed market price is taken to be a random sample of the random price; hence the correlation functions of market prices are taken to be equal to the average values of market prices, and which are empirically calculated by summing over the historical time series of market prices.

To empirically test and calibrate the statistical microeconomic formulation, the left hand side of Eq. \ref{microcorreprop} is computed using market data for prices. A best fit is then done for all the parameters of the model by using the right hand side of Eq. \ref{microcorreprop}.

Market prices are taken as independent inputs to the model and, from  Eq. \ref{minimamicropotenmodel}, are given by
\begin{eqnarray*}
p_{0i}=\left(\frac{a_id_i}{b_is_i}\right)^{1/(a_i+b_i)}
\end{eqnarray*}
Both indices $a_i$ and $b_i$ are taken as input, for example as obtained in \cite{gas1}.

Hence, market prices fix the ratios $d_i/s_i$.  The coefficient $\gamma_i$ and matrices of correlation $L_{ij}$ and $\bar{L}_{ij}$ are  fixed by empirically determining the propagator $G_{ij}(t-t')$. Empirically evaluating $\gamma_i^2=d_is_i$ then yields the values of $s_i$ and $d_i$.

Prices and quantities $\textbf{p}_0$ and $\textbf{q}_0$, similar to those given in Eq. \ref{pricquant}, can be derived for statistical economics and yield
\begin{eqnarray*}
\textbf{p}_0=\textbf{p}_0(\beta,\textbf{d})~~;~~\textbf{q}_0=\textbf{q}_0(\beta,\textbf{d})
\end{eqnarray*}
Taking the values $\textbf{p}_0$ and $\textbf{q}_0$ as \textit{input} and fixes the parameters $\beta_i$, which in turn, together with empirical values determined for $s_i$ yield the values for $\alpha_i$.

Hence all the parameters for the model can be empirically determined. Recall the  model for the demand and supply of commodity prices was fairly simple, given by Eq. \ref{micropotenmodel} as follows
\begin{eqnarray*}
\mathcal{V}[\textbf{p}]=\frac{m}{2}\left[\sum_{i=1}^N\frac{d_i}{p_i^{a_i}} + \sum_{i=1}^N s_ip_i^{b_i}\right]~;~~a,b,d_i,s_i>0 
\end{eqnarray*}
The simple form was chosen for the potential $\mathcal{V}[\textbf{p}]$ so that the analysis for computing the propagator as well as the general technique for calibrating this model could be carried out explicitly.

The precise form of the potential is not very important since, to leading order in $1/m$, the entire potential only contributes the parameter $\gamma_i$ to the value of the propagator. In fact, one can take the potential to be of the form
\begin{eqnarray*}
\mathcal{V}[\textbf{p}]=\frac{m}{2}\left[\sum_{i=1}^Nf(p_i) + \sum_{i=1}^N g(p_i)\right]
\end{eqnarray*}
Any empirical demand and supply function $f(p_i)$ and $g(p_i)$, respectively, that yield a unique set of market prices $\textbf{p}_0$ are equally good for modeling the microeconomic potential. Such a general potential would, to leading order in $1/m$, result in a different relation of $\gamma_i$ to the model's parameters, with all other results -- including the form of the propagator -- remaining unchanged.

\section{Summary}
A statistical generalization  of microeconomic modeling is proposed in this paper to consider all commodity prices to be stochastic processes.  The demand and supply function are interpreted as being components of a single underlying microeconomic potential and the average market price, to lowest order, is given by minimizing the microeconomic potential. A simple model for both the demand and supply functions have been proposed so that a concrete analysis could be carried out. The utility function was evaluated from the demand function using the principle of duality.

A Feynman path integral was defined for the random evolution of commodity prices and provides a theoretical framework for  the study of commodity prices considered as stochastic processes. 

The choice for the microeconomic kinetic term $\mathcal{T}[\textbf{p}]$ is based on a detailed empirical study of equity markets; the form chosen for $\mathcal{T}[\textbf{p}]$ has been shown by empirical evidence to be very accurate for a wide range of equities \cite{bebcyeqt}. The kinetic term driving the time dependence of commodity prices was proposed, in analogy with the behavior of equity prices, to be determined by the acceleration of commodity prices. This form of the kinetic energy leads to many new features not present in quantum mechanics. Furthermore, since commodities undergo a classical random evolution, many of the problems related to the lack of unitarity due to the acceleration term in the Lagrangian do not appear in microeconomics. 

The microeconomic potential term $\mathcal{V}[\textbf{p}]$ combines the demand and supply of commodities into a single entity and provides an entirely new perspective on the mode of competition between supply and demand. One needs to be studied for the major commodities; as mentioned earlier, and the empirical study of gasoline prices \cite{gas1} supports the form of the microeconomic potential chosen in this paper. 

The Lagrangian that combines the kinetic and potential terms for commodity prices shows the central role being played by the kinetic term; this term is absent in the standard treatments of microeconomic analysis that are focused almost solely on supply and demand. Of course, whether the kinetic term in fact is important in the dynamics of commodity prices is an empirical question and needs to be further studied. 

A well defined perturbation expansion about the minimum of the potential was defined and the propagator was explicitly  evaluated. The expansion of the path integral in terms of the inverse of the budget constraint is valid only for a large budget; if the budget becomes small, the statistical fluctuations become large and numerical methods are then necessary for evaluating the path integral. The expansion of market prices and its correlators in a power series in the inverse of the total budget, which has been introduced in this paper, needs to be studied empirically to ascertain whether in fact market data provides evidence of such an expansion.

The calibration and testing of the proposed statistical model of microeconomics is based on comparing the model's prediction with the empirical values of market prices as well as by  comparing the model's propagator (unequal time correlation function)  of market prices with the empirical propagator  obtained from market data. 
\section{Acknowledgment}
I am deeply indebted to Emmanuel Haven for having introduced me to the subject of Microeconomics,  and I thank him for many useful, stimulating and enjoyable discussions and for a careful reading of a draft of this paper. I thank Arzish Baaquie for a careful reading of the paper and making many valuable suggestions. I thank the School of Management, University of Leicester, for their warm hospitality during my sabbatical visit in 2011, and where the bulk of the work of this study was carried out.
\section{Appendix: Utility function}
The model considered in this paper starts with the demand function since the main focus is on market prices. For many theoretical studies in micro- and macro-economics the utility function plays a central role. A model for the utility function that could prove useful in such studies is the following
\begin{eqnarray}
\label{microutiltymodl}
\mathcal{U}&=&\frac{1}{2}\sum_{ij=1}^Nq_iM_{ij}q_j+\sum_{i=1}^Nh_iq_i~~;~q_i>0\\
           &\equiv&\frac{1}{2}qMq+hq \nonumber
\end{eqnarray}
where the last equation has been written in matrix notation.

Taking $M_{ij}, h_i>0$  fulfills the requirement for utility functions given in Section \ref{sec:utility}.

To obtain the demand function, the utility function is maximized with the constraint that the budget is fulfilled, namely
\begin{eqnarray}
\label{utilitydemandconst5}
\frac{\partial \mathcal{U}[\textbf{q}]}{\partial q_i}\Big{|}_{\textbf{q}=\bar{\textbf{q}}}=0~~;~~
\text{Constraint}:~\sum_{i=1}^Np_iq_i=m
\end{eqnarray}
Simultaneously solving the equations given in Eq. \ref{utilitydemandconst5} yields the value of $\bar{\textbf{q}}$ that maximizes the utility function for a given budget, namely
\begin{eqnarray*}
\bar{\textbf{q}}=\bar{\textbf{q}}(\textbf{p},m)~~\Rightarrow~~\mathcal{D}[\textbf{p},m]=\mathcal{U}[\bar{\textbf{q}}(\textbf{p},m)]
\end{eqnarray*}
Using the technique employed in Section \ref{sec:demandutily} to obtain the utility function from the demand function, it can be shown that, in matrix notation
\begin{eqnarray}
\label{demandutiaquad}
\bar{q}=M^{-1}(\zeta p-h)~~;~~\zeta=\frac{m+pM^{-1}h}{pM^{-1}p}\nonumber\\
\mathcal{D}[\textbf{p},m]=\frac{1}{2}\frac{(m+pM^{-1}h)^2}{pM^{-1}p}-\frac{1}{2}hM^{-1}h
\end{eqnarray}
The demand function derived in Eq. \ref{demandutiaquad} is not suitable for modeling the behavior of the market. When it is combined with the supply function to define the microeconomic potential $\mathcal{V}[\textbf{p},m]$, it can be shown that $\mathcal{D}[\textbf{p},m]$ given in in Eq. \ref{demandutiaquad} does not result in a unique minimum for the potential and hence does not yield a set of unique average market prices.
\bibliographystyle{is-unsrt}
\bibliography{../../master_references_all}
\end{document}